\documentclass[preprint,amsmath,amssymb]{revtex4}
\usepackage{graphicx}
\usepackage{dcolumn}
\usepackage{bm}
\begin{document}
\title{Casimir effect in a weak gravitational field and the spacetime index of refraction}
\author{ B.~Nazari\footnote{
Electronic address:~bornadel@khayam.ut.ac.ir} and M.~Nouri-Zonoz  \footnote{
Electronic address:~nouri@theory.ipm.ac.ir, corresponding author}}
\address{ Department of Physics, University of Tehran, North Karegar Ave., Tehran 14395-547, Iran.}
\begin{abstract}
In a recent paper \cite{0Nouri} using a conjecture it is shown how one can calculate the effect
of a weak stationary gravitational field on vacuum energy in the context of Casimir effect in an external gravitational field treated in $1+3$ formulation of spacetime decomposition.. 
In this article, employing quantum field theory in curved spacetime, we explicitly calculate the effect of a weak static gravitational field on virtual massless scalar particles in a Casimir apparatus. It is shown that, as expected from the proposed conjecture, both the frequency and renormalized energy of the virtual scalar field
are affected by the gravitational field through its index of refraction. This could be taken as a strong evidence in favour of the proposed conjecture. Generalizations to weak {\it stationary} spacetimes and virtual photons are also discussed.
\end{abstract}
\maketitle
\section{Introduction}
Zero point energy is a concept that its reality has always been the subject of somewhat intensive debate. Recently in the light of its important role in the formulation of the so called cosmological constant problem it has been discussed by people with opposite opinions \cite{1Jaffe,2Fulling}. In this same respect, one obvious
question need to be answered is its response to a gravitational field. In doing so in \cite{0Nouri} it is shown how combining the concept of a spacetime index of refraction (in $1+3$ formulation of spacetime decomposition) and virtual photon propagation in a Casimir apparatus could lead to a general solution in a weak stationary gravitational field in a coordinate-free manner. The conjecture employed there is  a simple one which assumes virtual photons are affected by the spacetime index of refraction in the same way that real photons do when propagate in a weak gravitational field.
It is shown that starting with this conjecture and assigning a frequency change $\omega \rightarrow \frac{\omega}{n}$ to virtual photon frequencies, one ends up with the same equation for the gravitational force on  Casimir energy as for any other mass/energy. In other words it is shown that the (finite) Casimir energy couples to all inertia like any other mass/energy. In what follows, using QFT in curved spacetime, as a first step we study Casimir effect for a massless scalar field in a weak static gravitational field to explicitly calculate both frequency and finite energy changes to show that it gives the same results conjectured in \cite{0Nouri}\footnote{This could be generalized to the case of a weak stationary field such as the Kerr spacetime in its weak field limit \cite{21Nazari}.}. This, in our opinion,
is a strong evidence that the proposed conjecture is correct. One should note that scalar field being massless, from  geometical point of view, moves on a null geodesic so it is expected that frequencies of the corresponding virtual massless particles to be affected in the same way that virtual photons do. On the other hand for the energy density of the scalar field, from the studies of QFT in curved spacetime and specially Casimir effect, it is well known \cite{3Birrell} that it should be one-half the electromagnetic case on account of the fact that electromagnetic field has twice as many modes. We will use this result to obtain the
electromagnetic Casimir energy in a weak gravitational field from that of a scalar field. In the last section we will present some general arguments as to why it is expected that the same method should in principle work for stationary spacetimes. 
\section{Fermi metric and weak static gravitational fields}
To solve the field equations for a massless scalar field in a weak gravitational field the first thing one should specify is the coordinates in which this is going to be done. On the other hand since the above mentioned conjecture is stated for weak gravitational fields, in this section we discuss in some detail our choice of coordinates for the weak field of a static gravitational field which is Fermi coordinates. In \cite{2Fulling} it is shown that coupling to gravity of Casimir finite energy for a weak gravitational field depends on the coordinate system employed. Analyzing this difficulty it is shown that it basiclly arises from the simple fact that relations between coordinate increments and physical distances depend upon the distance from the gravitating center in most common coordinate system. Then it is discussed why the calculations based on Fermi coordinates are the ones that should be taken as physical. The main reason being the fact that the Fermi coordinate is the closest generalization to a weak gravitational field of an inertial coordinate system in flat space \cite{30MTW, Misner, 5Marzlin}. So the metric of a weak static gravitational field, representing that of Earth or any other spherically symmetric mass, a distance $z$ above its surface of radius $R$, is chosen to be the Fermi metric
\begin{eqnarray}\label{metric1}
g_{00} = -(1 +  2{g} z)\;\;\;\; , \;\;\;\; g_{\mu\nu} = \delta_{\mu\nu}\;\;\;\;, \;\;\;\; \mu,\nu = 1,2,3
\end{eqnarray}
with ${g}=\frac{GM}{R^2}$ the gravitational acceleration. In brief this metric  represents the proper reference frame of an accelerated observer in a curved spacetime \footnote{Indeed Fermi coordinates in this case (i.e weak static gravitational field) incorporate only first derivatives of the metric such that the only non-zero component is $g_{00},i = -2a_i$ with  $a_i$ the observer's acceleration.}. In what follows we will employ the basic quantum field theoretic approach also used in \cite{Sorge} for the same problem but with the main difference that, in effect, the calculations are done in the weak field limit of Schwarzschild spacetime in isotropic coordinates. Of course by the above argument on the importance of the coordinates employed in such a calculation it is expected that our results to be at variance with those given in isotropic coordinates. We will compare the rersults and  elaborate on this important fact in the last section.
\section{Solution to the scalar field equation and mode frequencies}
Field equation for a massless scalar field $\Phi(x^\mu)$ in a curved background outside the matter distribution is given by
\begin{eqnarray}\label{FE}
\partial_a[\sqrt{-\mathfrak{g}} g^{ab} \partial_b \Phi(x^c)]= 0; \;\;\;\;
\mathfrak{g}\equiv {\rm det}g^{ab}; \;\;\;\; a,b = 0,1,2,3.
\end{eqnarray}
To quantize a scalar field in a curved spacetime one defines the following   generalization of a flat space scalar product \cite{3Birrell}
\begin{eqnarray}\label{eq001}
(\Phi_1 , \Phi_2) = -i\int_{\Sigma}\Phi_1(x)\partial_a \Phi_2^*(x)[-g_{\Sigma}(x)]^{\frac{1}{2}}n^a d\Sigma
\end{eqnarray}
in which $n^\mu$ is a unit timelike vector normal to the spacelike hypersurface $\Sigma$ (in our case $z$ = constant hypersurfaces ) and $d\Sigma$ is the volume element in $\Sigma$.  For a set of mode solutions of (\ref{FE}) orthonormal in the above product we have
\begin{eqnarray}\label{eq002}
(\Phi_i(x) , \Phi_j(x)) = \delta_{ij}\delta({\bf k}_i - {\bf k}_j)
\end{eqnarray}
Now for a weak static gravitational field represented by the metric (\ref{metric1}), the field equation (\ref{FE}) reduces to 
\begin{eqnarray}\label{FE1}
g^{00}\partial^2_t \Phi + \nabla^2 \Phi + {\bf\nabla} (ln\sqrt{-\mathfrak{g}}). {\bf\nabla}\Phi = 0
\end{eqnarray}
Our main goal here is to solve the above equation for modes inside the casimir apparatus. For simplicity we consider the case in which the plates are orthogonal to the radius along the $z$-axis in a local flat spatial metric of (\ref{metric1}) with its origin at the lower Casimir plate. Due to the static nature of the spacetime and the $x-y$ symmetry inside the cavity one could choose the scalar field to be of the following general (positive frequency) form
\begin{eqnarray}\label{SF0}
\Phi(x) = C e^{-i\omega t} e^{ik_x x}e^{ik_y y}Z(z)
\end{eqnarray}
in which $C$ is the normalization constant found through the orthonormality condition (\ref{eq002}) to be 
\begin{eqnarray}\label{C}
C^2 = \frac{1}{2(2\pi)^2 \omega} [\int_0^\ell \frac{Z^2(z)}{\sqrt{-g_{00}}} dz]^{-1}.
\end{eqnarray}
Substituting (\ref{SF0}) in equation (\ref{FE1}) leads to the following equation for the unknown function $Z(z)$,
\begin{eqnarray}\label{SF1}
Z''(z) + P(z) Z'(z) + Q(z) Z(z) = 0
\end{eqnarray}
in which $' \equiv \partial_z$, $k_\perp ^2=k_x^2 + k_y^2$ and
\begin{eqnarray}\label{eq1}
P(z) = (\partial_z ln\sqrt{-\mathfrak{g}}) \;\;\;\; ; \;\;\;\; Q(z) = -( k_\perp ^2 + \omega^2 g^{00}) \;\;\;\;
\end{eqnarray}
Now employing the common practice of writing $Z(z) = U(z) W(z)$ for the above linear, 2-nd order differential equation we end up with
\begin{eqnarray}\label{eq2}
W= exp(-\frac{1}{2}\int P(z) dz)= \frac{1}{(-\mathfrak{g})^{\frac{1}{4}}}\\
U'' + S(z) U =0\;\; ; \;\; S(z) = Q -\frac{1}{4}P^2 - \frac{1}{2} P'
\end{eqnarray}
Now it is well known that if $S' \ll 2|S|^{\frac{3}{2}}$ one could use the WKB approximation to solve the equation for $U$. It is an easy task to show that in our case (the weak static field) this condition is satisfied using the fact that $g=-(1 + 2gz)$ where
\begin{eqnarray}\label{eq02}
S = \frac{\omega^2}{-\mathfrak{g}} - k^2_\perp - \frac{3}{4}(\frac{g}{\mathfrak{g}})^2 \\ 
S^{\prime} = \frac{2g}{\mathfrak{g}^2} \omega^2 - 3(\frac{g}{\mathfrak{g}})^3
\end{eqnarray}
and the condition for the WKB approximation reduces to $g \ll \omega$ or equivalently $\frac{Gm\ell}{R^2} \ll 1$ which is obviously true. Therefore from WKB approximation we have the following general solution 
\begin{eqnarray}\label{eq03}
U = \frac{U_0}{S^{\frac{1}{4}}} \left(c_1 exp (i\int\sqrt{S(z)} dz) +  c_2 exp (-i\int\sqrt{S(z)} dz)\right) 
\end{eqnarray}
or appropriate for a Dirichlet boundary condition
\begin{eqnarray}\label{eq3}
U = \frac{U_0}{S^{\frac{1}{4}}} sin(A(z) +\phi_0)\\
A(z) = \int \sqrt{S(z)} dz
\end{eqnarray}
Applying the Dirichlet boundary condition to the above wavefunction on the plates i.e  $Z(0) = Z(\ell) =0 $ or equivalently $U(0) = U(\ell) =0 $ we have 
\begin{eqnarray}\label{eq4}
A(0) + \phi_0 = l\pi \\
A(\ell) + \phi_0 = m\pi
\end{eqnarray}
leading to the relation
\begin{eqnarray}\label{eq5}
A(\ell) - A(0) = \int_0^{\ell} \sqrt{S(z')} dz' = n\pi
\end{eqnarray}
in which $l,m,n$ are all integers. Now taking into account that in our case $g_{00}= 1+\epsilon$ with $\epsilon=2gz \ll 2 k_{\perp}$, from (\ref{eq02}) we have
\begin{eqnarray}\label{eq6}
S^{1/2} \approx (-\omega^2 g^{00} - k_{\perp}^2)^{1/2} 
\end{eqnarray}
which substituted in (\ref{eq5}) gives
\begin{eqnarray}\label{eq7}
\int_0^{\ell} (-\omega^2 g^{00} - k_{\perp}^2)^{1/2} dz' = n\pi
\end{eqnarray}
Working this integral out to the first order in $g\ell$ we find
\begin{eqnarray}\label{eq8}
\omega = (1+ \frac{g\ell}{2})\omega_0
\end{eqnarray}
with ${\omega_0}^2 = k_{\perp}^2 + (\frac{n\pi}{\ell})^2$ the field frequency in the absence of the gravitational field. The above relation is nothing but the proposed relation in \cite{0Nouri} for the frequency of the virtual photons in the cavity
\begin{eqnarray}\label{eq9}
\omega = \omega_0 (g_{00})^{\frac{1}{2}}= (1+2g<z>)^{\frac{1}{2}} \omega_0\approx (1+g <z> )\omega_0 
\end{eqnarray}
to the first order in $gz$. It should be noted that to compare the above two equations one has to take $ <z> = \frac{\int^{\ell}_0 z dz}{\ell}= \frac{\ell}{2}$, in other words the midway between $0$ and $\ell$ (average of $z$) on the Z-axis is taken to be the point where the index of refraction is calculated.\\
It should be noted that the above calculations are up to the first order in $g\ell$ (or $gz$) and this is so because the Fermi metric is the first order correction to the line element near an observer's worldline unaffected by the spacetime curvature. Spacetime curvature  comes into play only at the second order leading to the {\it Fermi normal coordinates} in the absence of acceleration and rotation \cite{Misner,30MTW}.
As pointed out in the introduction it should be noted that one should do the above calculation for a photon field to account for the electromagnetic Casimir effect. But since our scalar field is massless it also propagates on a null geodesic so it is expected that the same frequency relation holds for virtual scalar particles as it does for virtual photons. 
\section{Casimir energy in Fermi metric}
Employing the setting introduced in the begining of the last section the vacuum energy density for a scalar field is given by
\begin{eqnarray}\label{eq11}
\epsilon_{vac} = n^a n^b <0|T_{ab}|0>  
\end{eqnarray}
which in the present case reduces to
\begin{eqnarray}\label{eq11a}
\epsilon_{vac} = \frac{1}{-\mathfrak{g}} \Sigma_n \int d^2 k_{\perp}T_{00}[\Phi_n(k) , \Phi_n^*(k)]
\end{eqnarray}
where $T_{ab}[\phi_n(k) , \phi_n^*(k)]= \Phi_{,a} \Phi_{,b} - \frac{1}{2}g_{ab}g^{cd}\Phi_{,c} \Phi_{,d}$ is the bilinear expression for $T_{ab}$  \cite{3Birrell} reducing to the following expression for the $T_{00}$ component
\begin{eqnarray}\label{eq11b}
T_{00} = \frac{1}{2} C^2 Z^2(z)\left\lbrace \omega^2 - g_{00} (k_{\perp}^2 + (\frac{Z^\prime}{Z})^2\right\rbrace 
\end{eqnarray}
The cavity being in a curved background its energy density is position-dependent (in our case $z$-dependent) and so we consider the average Casimir energy density  given by
\begin{eqnarray}\label{eq11c}
\left\langle \epsilon_{vac}\right\rangle  = \frac{1}{\ell} \Sigma_n \int d^2 k_{\perp}\left\lbrace \int \frac{1}{-\mathfrak{g}} T_{00}[\Phi_n(k) , \Phi_n^*(k)]dz\right\rbrace 
\end{eqnarray}
which is a consequence of the fact that both the Fermi metric (\ref{metric1}) and $T_{00}$ are only z-dependent. We first calculate the integral inside the braces 
\begin{eqnarray}\label{eq12}
\int \frac{1}{-\mathfrak{g}} T_{00}[\Phi_n(k) , \Phi_n^*(k)]dz = \int (-g_{00})^{-1} \frac{1}{2} C^2 Z^2(z)[\omega^2 - g_{00}k_{\perp}^2 - g_{00}(\frac{Z^\prime}{Z})^2]dz
\end{eqnarray}
by approximating the function $Z(z)$
\begin{eqnarray}\label{eq13}
Z(z) \propto \frac{1}{(-\mathfrak{g}S(z))^{1/4}}{\rm sin} (\int \sqrt{S(z)} dz + \phi_0)
\end{eqnarray}
in which 
\begin{eqnarray}\label{eq14}
\int \sqrt{S(z)} dz = \int (\frac{\omega^2}{-g_{00}}-k_{\perp}^2 )^{\frac{1}{2}}dz \approx \int(\omega^2-k_{\perp}^2-2g\omega^2z)^{\frac{1}{2}}dz \cr
\equiv \int(az + b)^{\frac{1}{2}}dz \approx \sqrt{b}z 
\end{eqnarray}
where $b=\omega^2 - k_{\perp}^2$ and $a = -2g\omega^2$. In the same way
\begin{eqnarray}\label{eq15}
(-\mathfrak{g}S(z))^{1/4}= b^{\frac{1}{4}} (1+[\frac{g}{2} + \frac{a}{4b}]z)
\end{eqnarray}
So to the first order in $gz$ 
\begin{eqnarray}\label{eq16}
Z(z) \approx Z_0 (1-[\frac{g}{2} + \frac{a}{4b}]z) {\rm sin}(\sqrt{b} z)\;\;\;  {\rm and} \;\;\;\ \frac{Z^\prime}{Z}\approx \sqrt{b} cotg(\sqrt{b}z) - \frac{1}{2}(g+\frac{a}{2b})
\end{eqnarray}
Now substituting from (\ref{eq16}) into (\ref{eq12}) and after some manipulation we have
\begin{eqnarray}\label{eq17}
\int \frac {T_{00}}{-g_{00}}dz \approx \frac{1}{2} C^2 Z^2_0 \int(1-2gz)(1-(g+\frac{a}{2b})z) {\rm sin}^2(\sqrt{b} z)[\omega^2 - g_{00}(k_{\perp}^2 +(\frac{Z^\prime}{Z})^2]dz\cr
= \frac{1}{2} C^2 Z^2_0 \omega^2 \ell [1-(4g + \frac{a}{b})\frac{\ell}{4}]
\end{eqnarray}
On the other hand substituting for $Z(z)$ in (\ref{C}) from (\ref{eq16}) we have 
\begin{eqnarray}\label{eq18}
C^2 = \frac{1}{2(2\pi)^2\omega} \frac{2}{Z^2_0 \ell [1-(2g + \frac{a}{2b})\frac{\ell}{2}]}
\end{eqnarray}
So that (\ref{eq17}) reduces to
\begin{eqnarray}\label{eq18-a}
\int \frac {T_{00}}{-g_{00}}dz \approx \frac{\omega}{2(2\pi)^2}
\end{eqnarray}
Substituting this back into the equation for the average energy density we have
\begin{eqnarray}\label{eq19}
\left\langle \epsilon_{vac}\right\rangle  = \frac{1}{\ell} \Sigma_n \int d^2 k_{\perp}\left\lbrace \int \frac{1}{-g} T_{00} dz\right\rbrace \approx  \frac{1}{2(2\pi)^2 \ell}\Sigma_n \int d^2 k_{\perp}\omega
\end{eqnarray}
in which by substituting for $\omega$ from the previous section we end up with
\begin{eqnarray}\label{eq20}
\left\langle \epsilon_{vac}\right\rangle  = (1 +\frac{g\ell}{2})\left( \frac{1}{2(2\pi)^2\ell} \Sigma_n \int d^2 k_{\perp}\omega_0\right)
\end{eqnarray}
and finally employing the Schwinger proper-time method and any of the well known regularization methods to calculate the integral over the transverse momenta  and the infinite sum in the above expression respectively, we get
\begin{eqnarray}\label{eq20a}
{\left\langle \epsilon_{vac}\right\rangle}_{ren.}=(1 +\frac{g\ell}{2}) \epsilon^0_{vac}\equiv (1 +g<z>) \epsilon^0_{vac}  
\end{eqnarray}
In which $\epsilon^0_{vac}= -\frac{\pi^2}{1440\ell^4}$ is the renormalized vacuum energy density for a scalar field in a Casimir apparatus in flat spacetime \cite{3Birrell}. Although the electromagnetic case could be treated in the same way but as in the flat case it is expected that the electromagnetic value for the energy to be twice the above value for a scalar field on account of the fact that the electromagnetic field has twice as many modes i.e,
\begin{eqnarray}\label{eq21}
\left\langle \epsilon_{vac}\right\rangle_{EM} = (1 +\frac{g\ell}{2}) (2\epsilon^0_{vac})=(1 +\frac{g\ell}{2})(-\frac{\pi^2}{720\ell^4})\equiv (1 +{g<z>})(-\frac{\pi^2}{720\ell^4}).
\end{eqnarray}
Again this is the same result found in \cite{0Nouri} on the basis of the conjecture 
discussed in the introduction. It is also in agreement with the result found in  \cite{2Fulling}  using the gravitational energy formula for a static, weak field spacetime represented by the Fermi metric.
\section{discussion}
Using basic concepts of quantum field theory in a curved backround it is shown how the frequency and renormalized finite energy of virtual scalar particles in a Casimir apparatus in a weak static gravitational field are modified with respect to their flat space values. The calculations are done in a Fermi coordinate system which was shown to be the closest generalization to a weak gravitational field
of an inertial frame in flat space \cite{5Marzlin}.
As pointed out previously, the same calculations are also done in the weak field limit of Schwarzschild spacetime in isotropic coordinates \cite{Sorge}. On the other hand in \cite{2Fulling}, without employing quantum field theory in a curved background, it is shown that using the gravitational definition of the energy-momentum tensor for the electromagnetic Casimir stress tensor results in different values for the gravitational energy and force in different coordinate systems. In particular the values obtained for the gravitational energy and force in isotropic coordinates are shown to be twice those obtained in Fermi coordinates. Obviously there are no claculations for the photon frequencies in \cite{2Fulling} because of the non quantum field theoretric method used there, but surprisingly enough the frequency and energy modifications found in \cite{Sorge} (formulas (6.10) and (6.23)) for scalar particles in isotropic coordinates are twice those obtained in the present article \footnote{Indeed the sign of the value for energy in \cite{Sorge} is different from our result and that in \cite{2Fulling} which leads to a lower absolute value for energy compared to the flat case inspite of the fact that the fequencies are increased.}. This is expected from the arguments about the relevance of the coordinate system used to represent the weak field limit in \cite{2Fulling}. It should be noted that the value obtained for the finite Casimir energy in the present article was a direct consequence of the value found for the particle frequencies.
Generalising our field theorertic approach to a stationary spacetime one could go back to the $1+3$ formulation of stationary spacetimes in which the spacetime line element is written in the following form
\begin{eqnarray}\label{metric}
ds^2 = g_{00}(dx^0 -{A_\alpha}dx^\alpha)^2 -dl^2
\end{eqnarray}
where $A_{\alpha}=-\frac{g_{0\alpha}}{g_{00}}$ is the so called gravitomagnetic potential and 
\begin{eqnarray}
dl^2 = \gamma_{\alpha\beta}dx^{\alpha}dx^{\beta} = (-g_{\alpha\beta} + \frac{g_{0\alpha}g_{0\beta}}{g_{00}})dx^\alpha dx^\beta\;\;\;\alpha, \beta = 1,2,3
\end{eqnarray}
is the spatial length element in terms of the three-dimensional spatial metric $\gamma_{\alpha\beta}$. In the weak field and slow orbit (${\rm v} \ll c$) limit the spacetime metric could be written in the following form \cite{Rindler},
\begin{eqnarray}\label{eq22}
ds^2 \approx (1 + \frac{2\Phi}{c^2} - \frac{2 {\bf A}. {\bf v}}{c^2})c^2 dt^2 - dl^2
\end{eqnarray}
In which $e^{\frac{2\Phi}{c^2}} = g_{00}$ and ${\rm v}^\alpha = \frac {dx^{\alpha}}{dt}$.
In some sense in the metric form (\ref{eq22}) the cross term has been absorbed into the time-time component. In other words, in these limits, it looks as if one could think of a stationary spacetime as a static one with a more complicated $g_{00}$ component leading to the correct form for the spacetime index of refraction ($n = \frac {1}{\sqrt{g_{00}}}\approx 1-\Phi + {\bf A}. {\bf v}; \;\;\; c=1$).\\
Generalizing the results of the previous sections to the case of a weak stationary spacetime (say of a rotating spherical mass) the natural approach would be the application of the quantum field theoretic methods to the case of a Casimir apparatus in a spacetime represented by a Fermi coordinate system which includes rotation \cite{30MTW}
\begin{eqnarray}\label{eq24}
ds^2 = (1 + 2a_\mu x^\mu)dt^2 + 2(A_\mu dx^\mu) dt - \delta_{\mu\nu} dx^\mu dx^\nu
\end{eqnarray}
in which $A_\mu = \epsilon_{\mu \nu \beta}x^\nu \omega^\beta$ and $\omega^\beta$ is the observer's angular velocity relative to inertial-guidance gyroscopes.
But from the above argument one could transform (\ref{eq24}) to the following equivalent form
\begin{eqnarray}\label{eq25}
ds^2 = (1 + 2a_\mu x^\mu +2A_\mu {\rm v}^\mu )dt^2  - \delta_{\mu\nu} dx^\mu dx^\nu
\end{eqnarray}
Comparing this metric with equation (\ref{metric1}) it suggests that the previous results for a static spacetime will be repeated but now with the above modified $g_{00}$ component. Of course this should be compared with the result obtained from the direct application of quantum field theory in a background spacetime represented by metric (\ref{eq24}) \cite{21Nazari}.

\section *{Acknowledgments}
The authors would like to thank the University of Tehran for supporting this project under the grants provided by the research council. M. N-Z also thanks the Center of Excellence on the Structure of Matter of the Unversity of Tehran.
\pagebreak

\end{document}